\documentclass[12pt]{iopart}
\usepackage{graphicx}
\usepackage{epsfig}
\usepackage{pennames}

\newcommand{\Aref}[1]{$^{#1}$}
\newcommand{\mT}{\ensuremath{m_{\mathrm T}}}
\newcommand{\pT}{\ensuremath{p_{\mathrm T}}}

\newcommand{\dder}{\ensuremath{{\mathrm d}}}
\newcommand{\ycm}{\ensuremath{y_{\mathrm{cm}}}}
\begin{document}

\title[]{Strangeness enhancements at central rapidity\\
         in 40$\:A$ GeV/$c$ Pb--Pb collisions}

\author{
  NA57 collaboration \\[2ex]
  F~Antinori\Aref{13},
  P~A~Bacon\Aref{6},
  A~Badal\`{a}\Aref{7},
  R~Barbera\Aref{8},
  A~Belogianni\Aref{1},
  I~J~Bloodworth\Aref{6}$^,$\Aref{23},
  M~Bombara\Aref{11},
  G~E~Bruno\Aref{2},
  S~A~Bull\Aref{6},
  R~Caliandro\Aref{2},
  M~Campbell\Aref{9},
  W~Carena\Aref{9},
  N~Carrer\Aref{9},
  R~F~Clarke\Aref{6},
  A~Dainese\Aref{13},
  D~Di~Bari\Aref{2},
  S~Di~Liberto\Aref{18},
  R~Divia\Aref{9},
  D~Elia\Aref{3},
  D~Evans\Aref{6},
  G~A~Feofilov\Aref{20},
  R~A~Fini\Aref{3},
  P~Ganoti\Aref{1},
  B~Ghidini\Aref{2},
  G~Grella\Aref{19},
  H~Helstrup\Aref{5},
  K~F~Hetland\Aref{5},
  A~K~Holme\Aref{12},
  A~Jacholkowski\Aref{7},
  G~T~Jones\Aref{6},
  P~Jovanovic\Aref{6},
  A~Jusko\Aref{6},
  R~Kamermans\Aref{22},
  J~B~Kinson\Aref{6},
  K~Knudson\Aref{9},
  V~Kondratiev\Aref{20},
  I~Kr\'alik\Aref{10},
  A~Krav\v{c}\'{a}kov\'{a}\Aref{11},
  P~Kuijer\Aref{22},
  V~Lenti\Aref{3},
  R~Lietava\Aref{6},
  G~L\o vh\o iden\Aref{12},
  V~Manzari\Aref{3}$^,$\Aref{9},
  M~A~Mazzoni\Aref{18},
  F~Meddi\Aref{17},
  A~Michalon\Aref{21},
  M~Morando\Aref{14},
  P~I~Norman\Aref{6},
  A~Palmeri\Aref{7},
  G~S~Pappalardo\Aref{7},
  R~J~Platt\Aref{6},
  E~Quercigh\Aref{9},
  F~Riggi\Aref{8},
  D~R\"{o}hrich\Aref{4},
  R~Romita\Aref{2},
  K~\v{S}afa\v{r}\'{\i}k\Aref{9},
  L~\v{S}\'andor\Aref{10},
  E~Schillings\Aref{22},
  G~Segato\Aref{14},
  M~Sen\'{e}\Aref{15},
  R~Sen\'{e}\Aref{15},
  W~Snoeys\Aref{9},
  F~Soramel\Aref{14},
  M~Spyropoulou-Stassinaki\Aref{1},
  P~Staroba\Aref{16},
  R~Turrisi\Aref{13},
  T~S~Tveter\Aref{12},
  J~Urb\'{a}n\Aref{11},
  P~van~de~Ven\Aref{22},
  P~Vande~Vyvre\Aref{9},
  A~Vascotto\Aref{9},
  T~Vik\Aref{12},
  O~Villalobos Baillie\Aref{6},
  L~Vinogradov\Aref{20},
  T~Virgili\Aref{19},
  M~F~Votruba\Aref{6},
  J~Vrl\'{a}kov\'{a}\Aref{11} and
  P~Z\'{a}vada\Aref{16} \\[2ex]
}

\address{\scriptsize
  $^{1}$  University of Athens, Athens, Greece \\
  $^{2}$  Dipartimento Interateneo di Fisica ``M. Merlin'' and Sezione INFN,
          Bari, Italy \\   
  $^{3}$  Sezione INFN, Bari, Italy \\
  $^{4}$  Fysisk institutt, Universitetet i Bergen, Bergen, Norway \\
  $^{5}$  H\o gskolen i Bergen, Bergen, Norway \\
  $^{6}$  University of Birmingham, Birmingham, UK \\
  $^{7}$  Sezione INFN, Catania, Italy \\
  $^{8}$  Dipartimento di Fisica dell'Universit\`{a} di Catania and
          Sezione INFN, Catania, Italy \\
  $^{9}$  CERN, European Laboratory for Particle Physics, Geneva, Switzerland \\
  $^{10}$ Institute of Experimental Physics, Slovak Academy of Science,
          Ko\v{s}ice, Slovakia \\
  $^{11}$ P.J. \v{S}af\'{a}rik University, Ko\v{s}ice, Slovakia \\
  $^{12}$ Fysisk institutt, Universitetet i Oslo, Oslo, Norway \\
  $^{13}$ Sezione INFN, Padova, Italy \\
  $^{14}$ Universit\`{a} di Padova and Sezione INFN, Padova, Italy \\
  $^{15}$ Coll\`{e}ge de France, Paris, France \\
  $^{16}$ Institute of Physics, Prague, Czech Republic \\
  $^{17}$ Sapienza Universit\`{a} di Roma and Sezione INFN di Roma1, Rome, Italy \\
  $^{18}$ Sezione INFN di Roma1, Rome, Italy \\
  $^{19}$ Universit\`{a} di Salerno and Sezione INFN, Salerno, Italy \\
  $^{20}$ University of St. Petersburg, St. Petersburg, Russia \\
  $^{21}$ IReS/ULP, Strasbourg, France \\
  $^{22}$ Utrecht University and NIKHEF, Utrecht, The Netherlands \\
  $^{23}$ Deceased.
}

\ead{ladislav.sandor@cern.ch}
\begin{abstract}
Results are presented on K$^0_\mathrm{S}$, hyperon and antihyperon
production in \mbox{Pb--Pb} and p--Be interactions at 40 GeV/$c$ per nucleon.
 The enhancement pattern follows the same hierarchy as seen in the higher energy data - the enhancement increases with the strangeness content of the hyperons and with the centrality of collision. The centrality dependence of the Pb--Pb yields and enhancements is steeper at 40 than at 158 $A$ GeV/$c$. The energy dependence of strangeness enhancements at mid-rapidity is discussed.

\end{abstract}


\section{Introduction}

Results are presented on the production of K$^0_\mathrm{S}$ mesons and $\Lambda, \Xi$ and $\Omega$ hyperons in \mbox{Pb--Pb} collisions at 40 $A$ GeV/$c$ beam momentum, and compared with p--Be reference data collected at the same momentum.

The data were obtained by the NA57 experiment at the CERN SPS. Its aim was to extend the study of strange- and multistrange-particle enhancements in heavy-ion collisions, initiated by the CERN WA97 experiment \cite{wa97_1,wa97_2}, by investigating the dependence of the enhancements on beam energy and on collision centrality. Several results from the NA57 experiment
on the production of strangeness have already been published. Below, we briefly summarize the most important ones:

\begin{itemize}
 \item The enhancement pattern of strange-hadron abundances in central Pb--Pb collisions, first observed by  WA97 at 158 $A$ GeV/$c$, has been confirmed.
 The enhancement grows with the strangeness of the produced hadrons and amounts to about a factor of 20 for $\Omega$ for the most central collisions \cite{enh06}. Such a pattern had been predicted to be one of  the signatures of the transition between hadronic matter and a plasma of quarks and gluons \cite{RaMu,KoMuRa}. The enhancements increase with the number of nucleons participating in the collision across the whole explored centrality range, from rather peripheral (about 50 participant nucleons) to central collisions. So far, it has not been possible to reproduce these results within conventional hadronic models of nucleus--nucleus collisions.
  \item A blast-wave model analysis of the transverse mass distributions of strange particles
\cite{na57_mt1,na57_mt2} indicates a collective expansion at a speed close to 50\% of the light velocity, superimposed on a thermal distribution. This suggests the generation of a very strong pressure, as would be expected in the case of deconfinement, owing to the large increase in the number of degrees of freedom. The study of the hydrodynamical conditions for semi-central collisions is of particular interest due to the observation, at the SPS, of significant azimuthal anisotropy in the momentum distribution of produced particles. At RHIC, the observation of an azimuthal anisotropy near the hydrodynamical limit in semi-central collisions constituted one of the main pillars for the `perfect liquid' claim
(see e.g., \cite{brahms,phobos,star,phenix}). At the SPS, a substantial increase of the freeze-out temperature was observed for more peripheral collisions. 
The NA57 measurement of the centrality dependence of the freeze-out temperature can be used to calculate the hydrodynamic limits for comparison with the $v_{\mathrm{2}}$ measurements at the SPS \cite{v2_na45,v2_na49}. To our knowledge this has yet to be done.
 \item The $\pT$ dependence of the nuclear modification factor $R_{\mathrm{CP}}$ obtained from the analysis of the
centrality dependence of  K$^0_\mathrm{S}$ and $\Lambda$ production at 158 $A$ GeV/$c$
\cite{rcp}, shows a pattern similar to that observed at RHIC, which has been interpreted as due to energy loss in the medium formed in the collision.
\end{itemize}

The emerging picture, also considering that the values of the hyperon enhancements measured at RHIC \cite{enh_star} are similar to those measured at the SPS, is that there does not seem to be a qualitative difference between the mechanisms at work in strangeness production in the systems created in heavy-ion collisions at the two energies.

Preliminary results on the strangeness enhancement at 40 $A$ GeV/$c$ beam momentum have also been published (see e.g., \cite{prel40}). They will be completed and discussed in the present paper which is organized as follows. The NA57 apparatus, as used for the Pb--Pb and p--Be data taking, is described in section 2. In section 3 the extraction of samples of strange particles and the methods used to correct the data for experimental biases is explained. In section 4 the measured yields and enhancements as functions of the centrality are shown. The energy dependence of strange particle yields and enhancements is presented in section 5. Finally, section 6 contains discussion and conclusions.
 
\section{The NA57 experiment} 
The NA57 experiment at the CERN SPS was a dedicated experiment for the study of the strange and multistrange particle
production in Pb--Pb and \mbox{p--Be} collisions \cite{na57}. It continued and extended the study performed by its predecessor WA97 by enlarging the available centrality region and collecting data at two beam momenta -- 158 and 40 $A$ GeV/$c$.

The layout of the NA57 experiment was conceptually similar to that of WA97 (for details see \cite{enh06,manzari1,manzari2,virgili}).
The $\PgL, \,\PgXm, \,\PgOm$ hyperons, their anti-particles and the 
\mbox{$\PKzS$~mesons} were identified by reconstructing their weak decays into final states containing only charged particles: $\PgL \rightarrow \Pp + \Pgpm$,
$\PgXm \!\! \rightarrow \PgL + \Pgpm$, $\PgOm \!\! \rightarrow \PgL + \PKm$, 
$\PKzS \rightarrow \Pgpm + \Pgpp$ plus the corresponding charge conjugate decays for anti-hyperons. The charged tracks emerging from strange particle decays were reconstructed in a telescope composed of an array of thirteen silicon pixel detector planes of 
$\mathrm{5} \! \times \! \mathrm{5~cm}^2$ cross-section placed in an approximately uniform 1.4~T magnetic field perpendicular to the beam line in the GOLIATH magnet. Most of the pixel planes were closely packed in an approximately 30 cm long compact part used for track finding.
The telescope was placed above the beam line, inclined, with the lower edge of the detectors lying on a line pointing to the target. The inclination angle $\alpha$ and the distance $d$ of the first plane from the target were set so as to accept particles produced in about a unit of rapidity around mid-rapidity, with transverse momentum above a few hundred MeV/$c$ (see section 3). Four additional pixel detector planes, together with four planes of double-sided silicon microstrip detectors, formed a lever arm system placed downstream in the magnetic field to enhance the momentum resolution for high momentum tracks. In the following, we describe the layouts used for collecting the Pb--Pb and p--Be events at 40 $A$ GeV/$c$.

\begin{figure}[t!]
 \centering
 \resizebox{0.85\textwidth}{!}{%
 \includegraphics{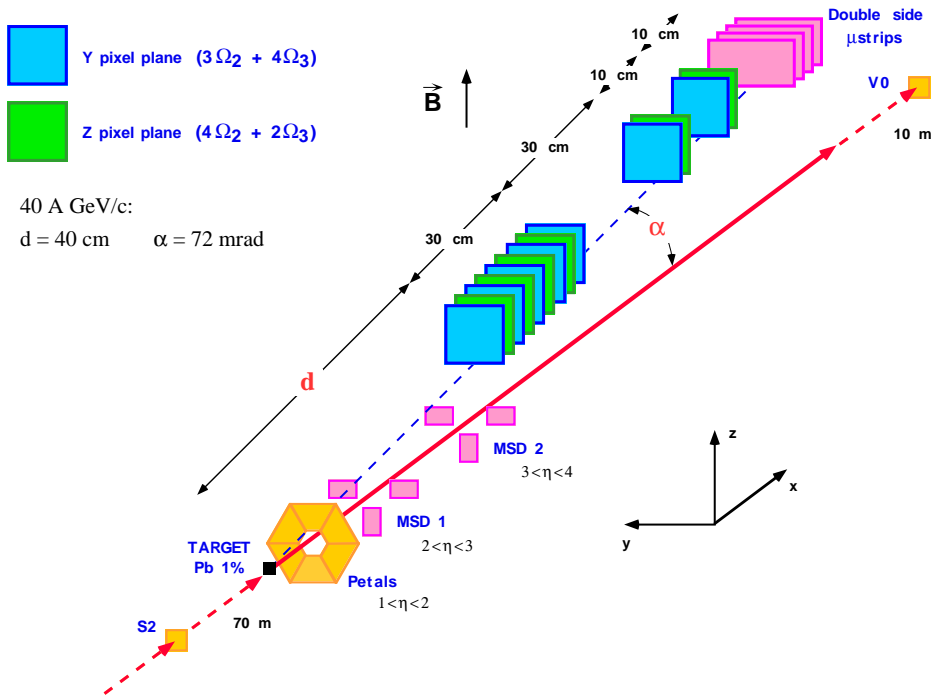}}
\caption {\label{setup_pbpb} The configuration of NA57 apparatus used  for the study of Pb--Pb collisions at 40 $A$ GeV/$c$.}
\end{figure}

The experimental set-up for Pb--Pb collisions is shown in figure \ref{setup_pbpb}. The thickness of the Pb target was 1\% of an interaction length for a lead beam. The inclination angle of the telescope was 72 mrad and the distance of the first plane from the target was \mbox{40~cm}. The hexagonal array of six scintillation counters (so called `petals'), placed 7.9 cm downstream of the target, provided a fast signal for triggering on the collision centrality. The petals covered the pseudorapidity
interval 0.8 $< \eta <$ 1.8 and the thresholds of the counters were tuned
so that the triggered-event sample corresponds to the 56\% most central inelastic Pb--Pb cross-section. 

The centrality of the Pb--Pb collisions was determined offline by analysing the charged-particle multiplicity measured by two stations of silicon microstrip detectors (MSD) sampling the pseudorapidity intervals 1.9 $< \eta <$ 3.0 and 2.4 $< \eta <$ 3.6 . 
  
The apparatus was exposed to a 40 $A$ GeV/$c$ Pb beam extracted from the CERN SPS and 240 million events were collected. The magnetic field polarity was periodically inverted during both Pb--Pb and p--Be data taking.  

\begin{figure}[t!]
\centering
 \resizebox{0.70\textwidth}{!}{%
 \includegraphics{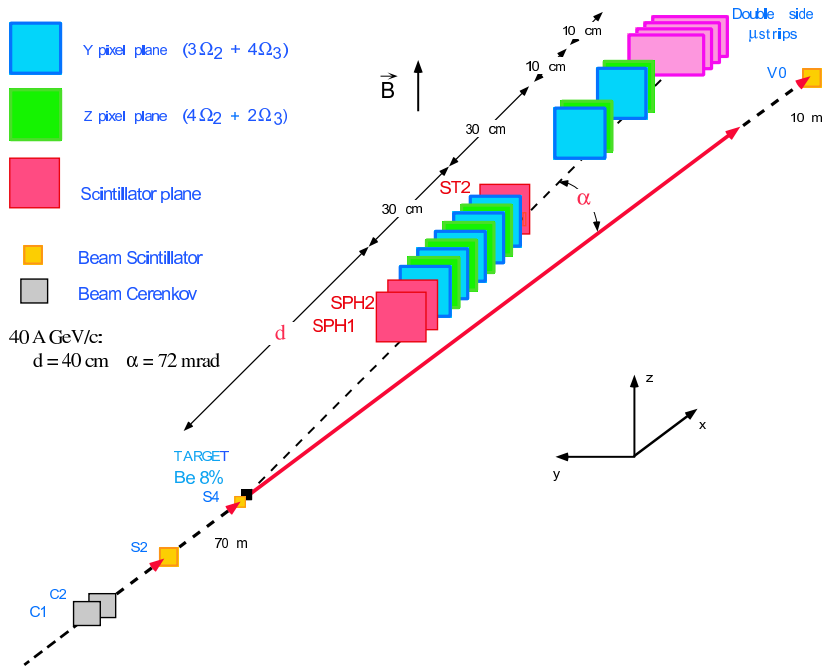}}
\caption {\label{setup_pbe} The layout of the NA57 set-up for the 
   study of 40 GeV/$c$ p--Be interactions.}
\end{figure}

The configuration of the NA57 apparatus used to collect the p--Be reference data is shown in figure \ref{setup_pbe}. It differs from the Pb--Pb set-up mainly in the triggering detectors.

The beam protons were selected using gaseous Cherenkov threshold counters C1 and C2 placed in the beam line upstream of the target. Three scintillation counters SPH1, SPH2 and ST2 were used in the trigger to enhance the probability to detect hyperon decay tracks in the telescope. The scintillators were placed upstream (SPH1 and SPH2) and downstream (ST2) of the compact part of the telescope. Their purpose was to select events where at least two charged particles passed through the telescope.

The beryllium target was 3.26 cm thick, which corresponds to 8\% of the p--Be interaction length. 

This experimental set-up was exposed to a 40 GeV/$c$ secondary proton beam at the
CERN SPS in two separate running periods and 170 million p--Be events were collected in total. 

\section{Data samples and analysis}

\begin{figure}[t]
\centering
\resizebox{0.50\textwidth}{!}{%
\includegraphics{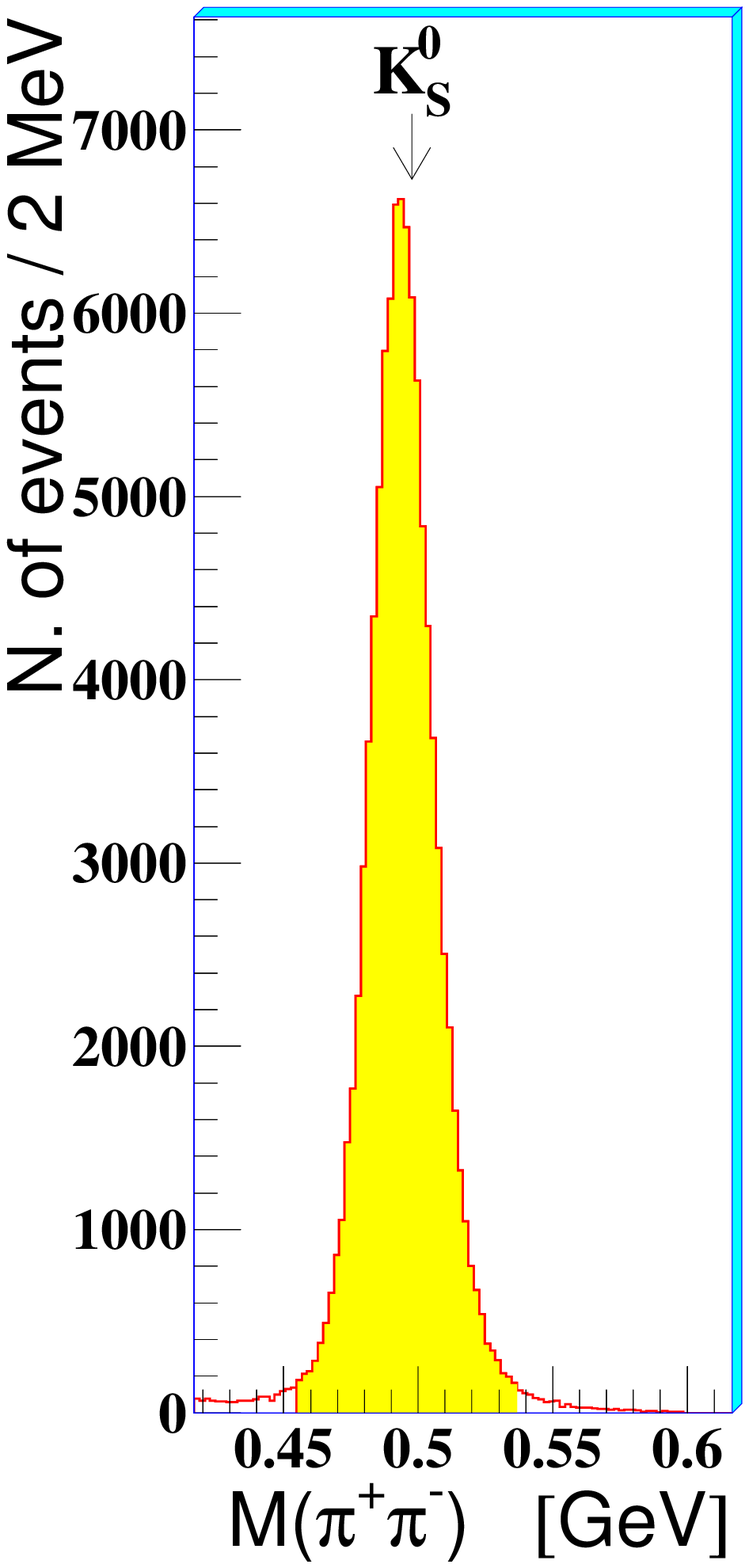}
\hspace{-0.5cm}
\includegraphics{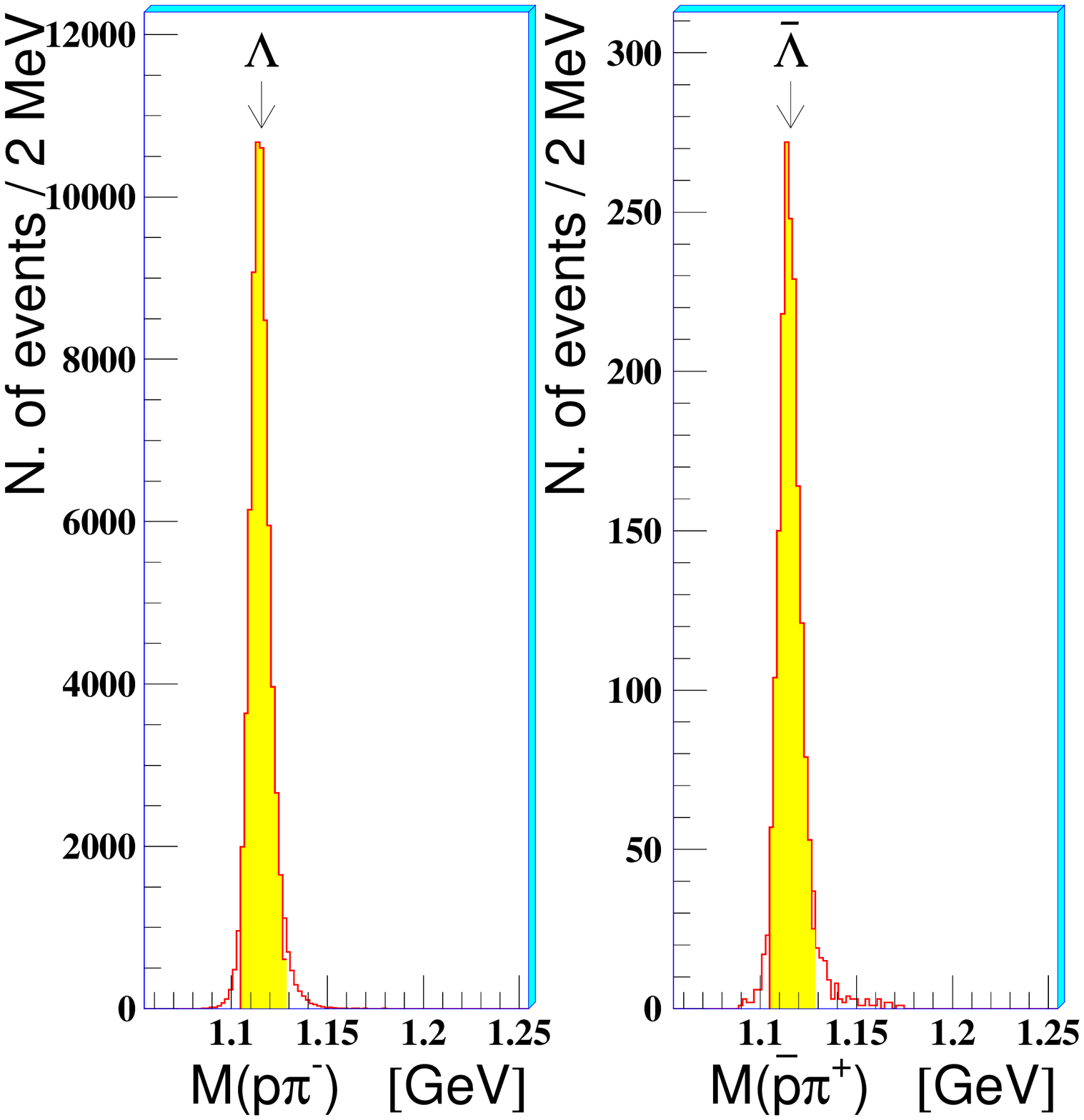}} \vspace{-0.25cm}
\resizebox{0.65\textwidth}{!}{%
\includegraphics{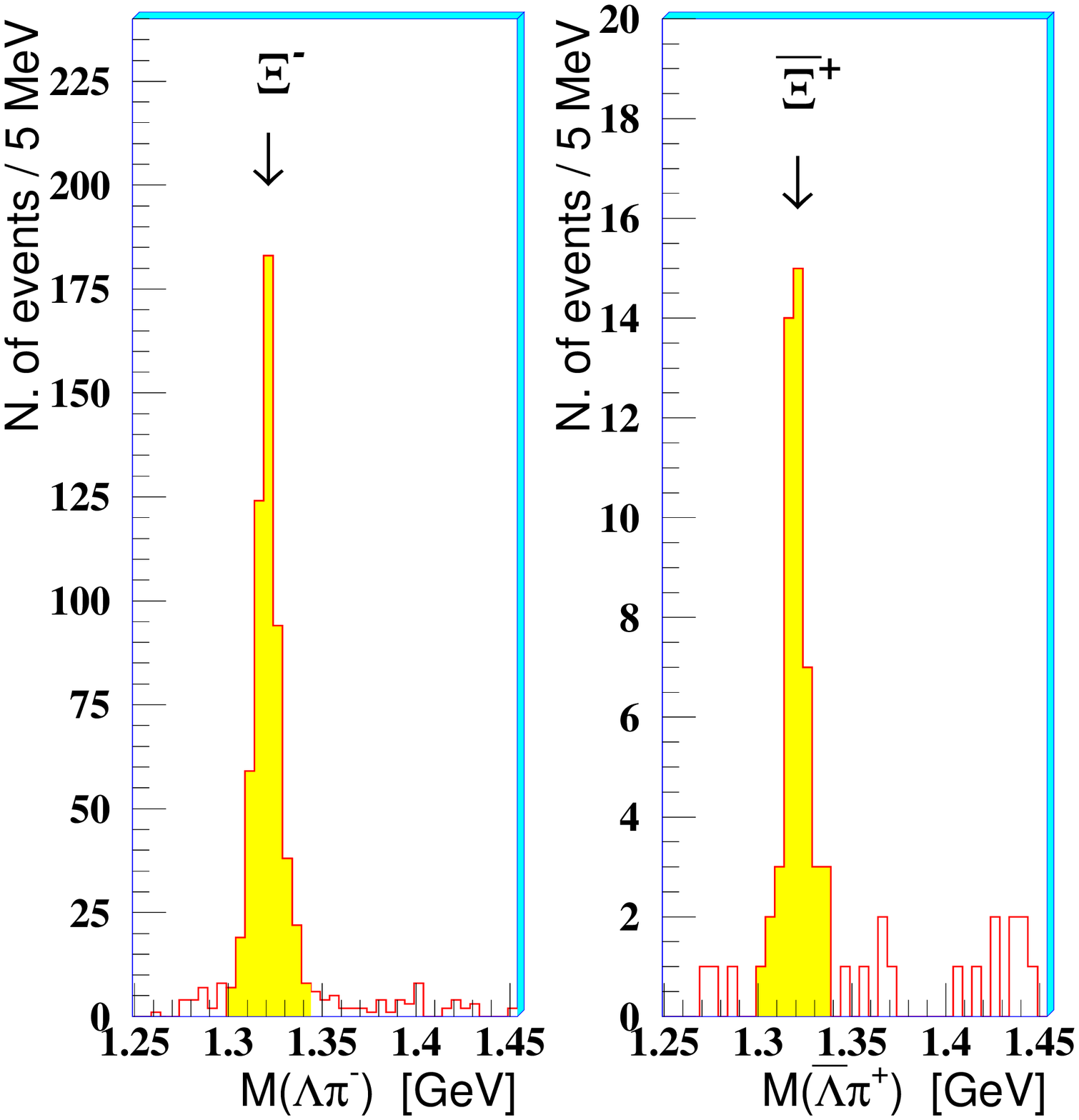}
\hspace{-1.5cm}
\includegraphics{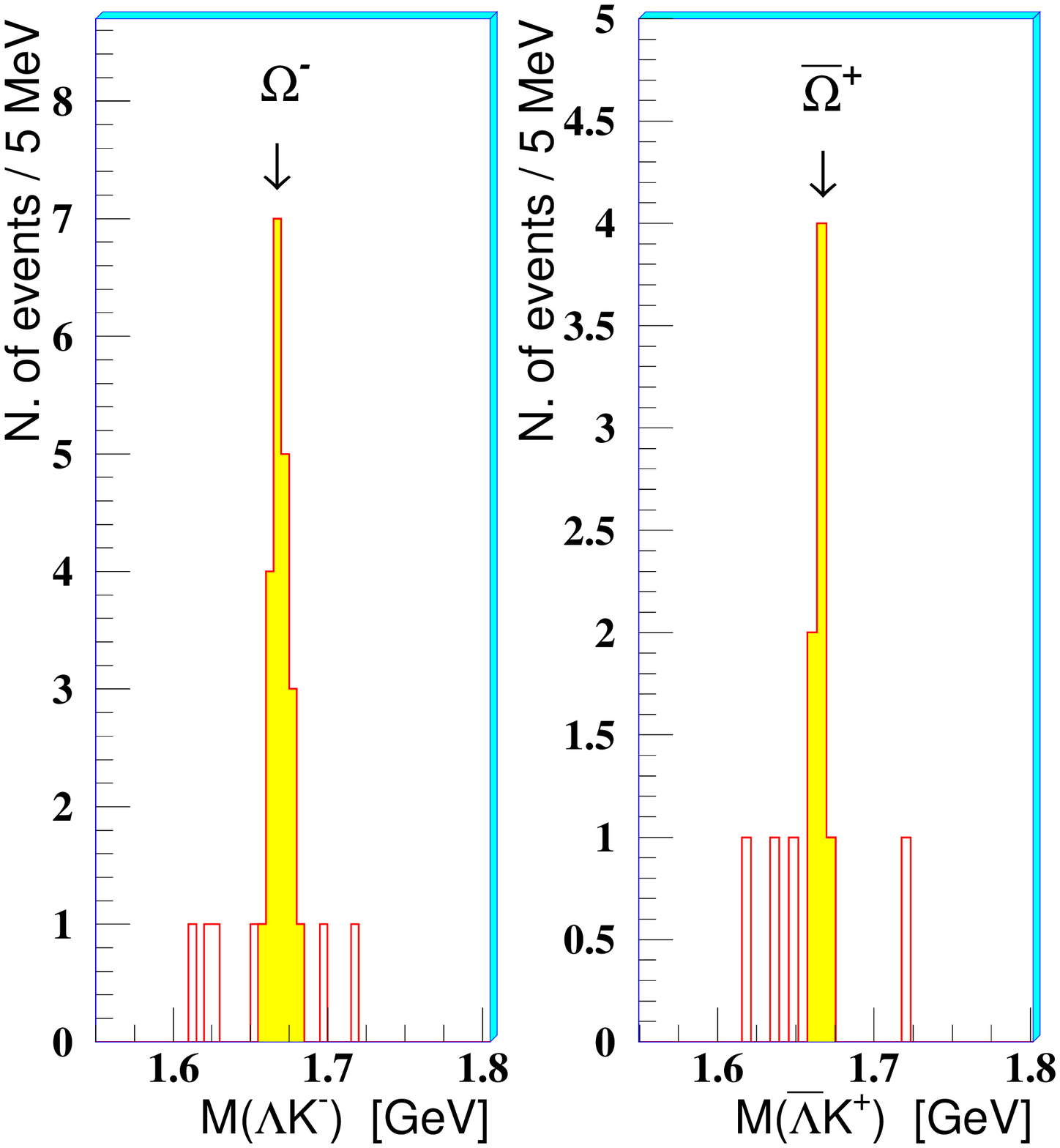}}
\caption{Invariant mass spectra for strange particle samples in Pb-Pb collisions.}  
\label{fig:masses_pb}
\end{figure}

The signals of $\PKzS$ mesons, hyperons and anti-hyperons were extracted using geometrical and kinematical constraints, following a procedure similar to that used for the \mbox{158~$A$~GeV/$c$} data (see \cite{enh06} and references therein). The selection criteria used in the \mbox{Pb--Pb} and p--Be data analysis allowed us 
to remove a substantial amount of combinatorial background. This is illustrated by the invariant mass spectra shown in figure \ref{fig:masses_pb} for Pb--Pb interactions with all cuts except the one on the invariant mass applied. The hyperon invariant mass spectra peak at the nominal PDG values \cite{pdg} with FWHMs of about 15 MeV/$c^2$. The $\PKzS$ mass peak has a FWHM of 25 MeV/$c^2$ and its maximum is lower than the nominal mass value by about 4 MeV/$c^2$.

\begin{figure}[b]
\centering
\resizebox{0.73\textwidth}{!}{%
\includegraphics{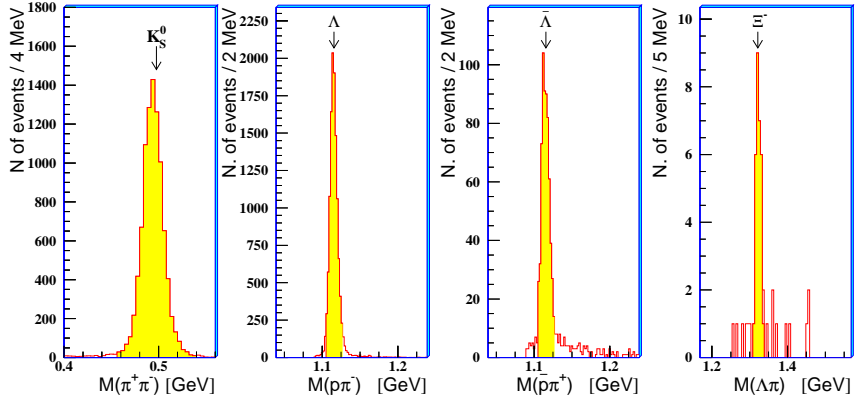}}
\caption{Invariant mass spectra for strange particles in p-Be collisions.}  
\label{fig:masses_be}
\end{figure}

The amount of remaining combinatorial background has been estimated by the event mixing technique \cite{na57_mt1}. The residual background for cascade hyperons has been evaluated to be about 4\% for $\PgXm$ and less than 10\% for $\PagXp$ and $\PgOm+\PagOp$; for singly strange particles the background has been estimated to be 1\%, 0.8\% and 2\% for $\PKzS,\PgL$ and $\PagL$, respectively. These backgrounds could influence the shape of transverse mass spectra and have been accounted for in the evaluation of the systematic errors.

The corresponding invariant mass plots for the p--Be data are shown in figure \ref{fig:masses_be}. The strange particle peaks are again well centred with FWHMs similar to those of the Pb--Pb samples except for the $\PKzS$ mass peak, which similarly to the Pb--Pb case is shifted down by \mbox{3.5~MeV/$c^2$} with respect to the PDG value.

\begin{figure}[b]
\centering
\resizebox{0.6\textwidth}{!}{%
\includegraphics{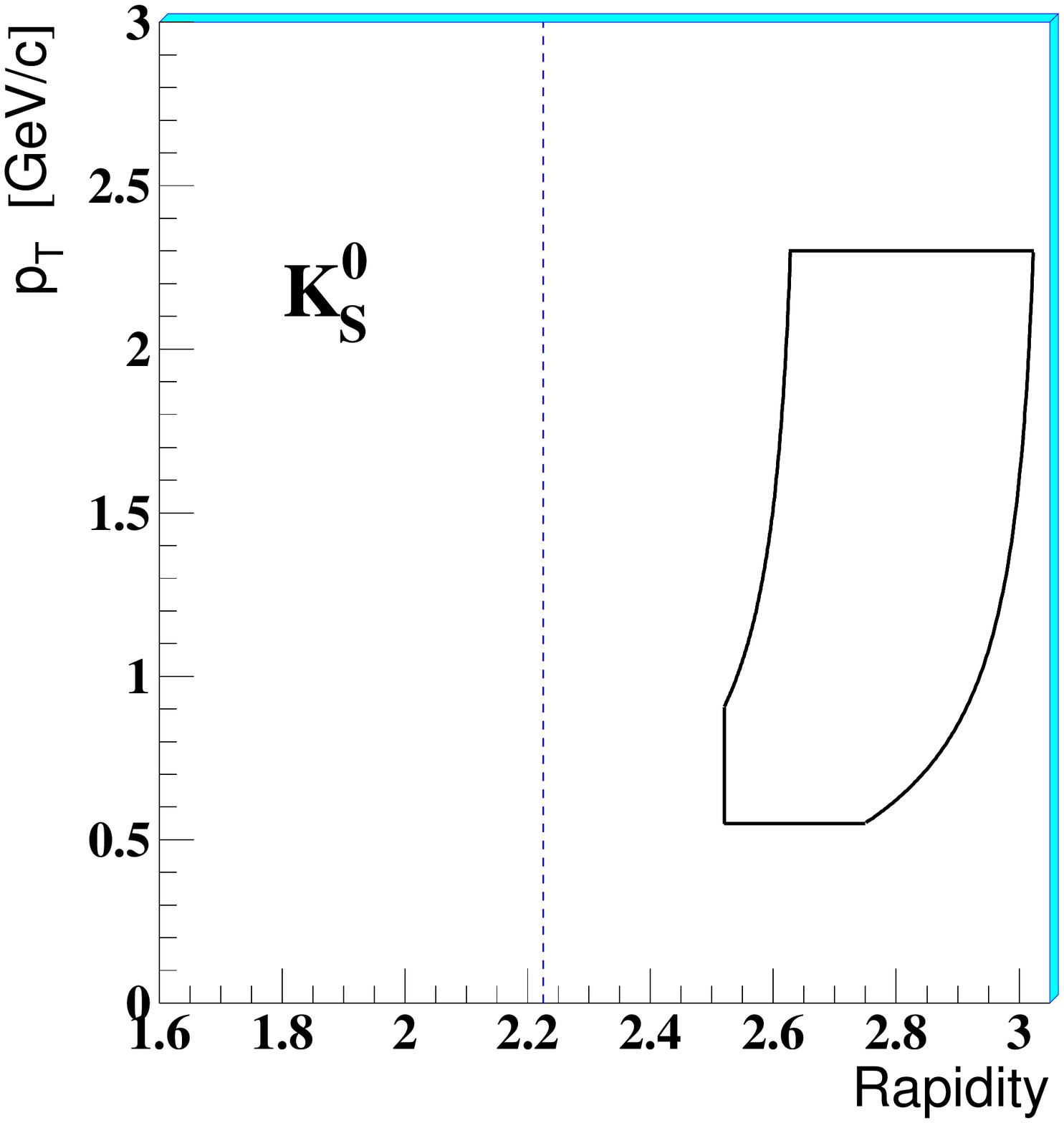}
\includegraphics{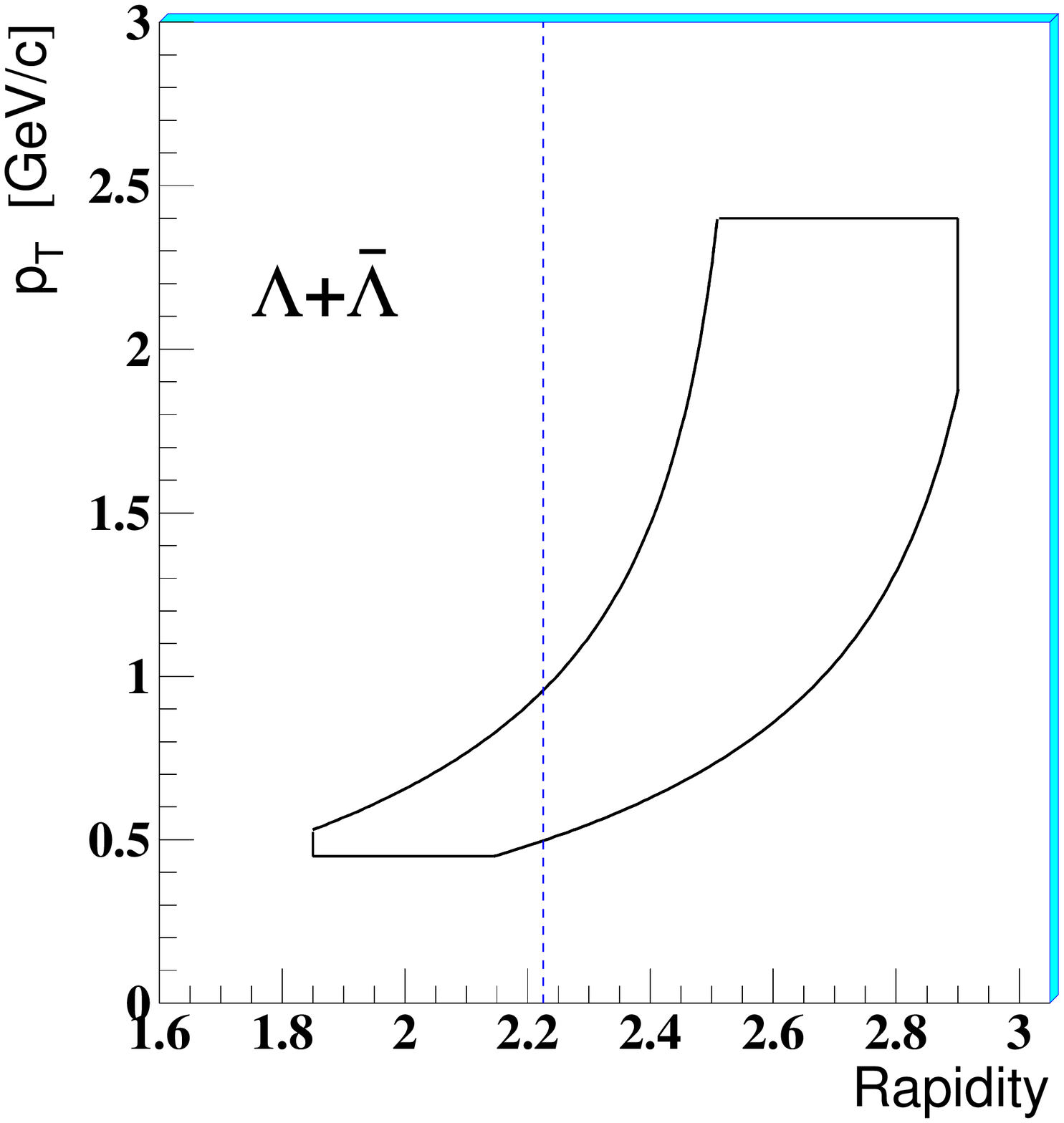}}
\resizebox{0.6\textwidth}{!}{%
\includegraphics{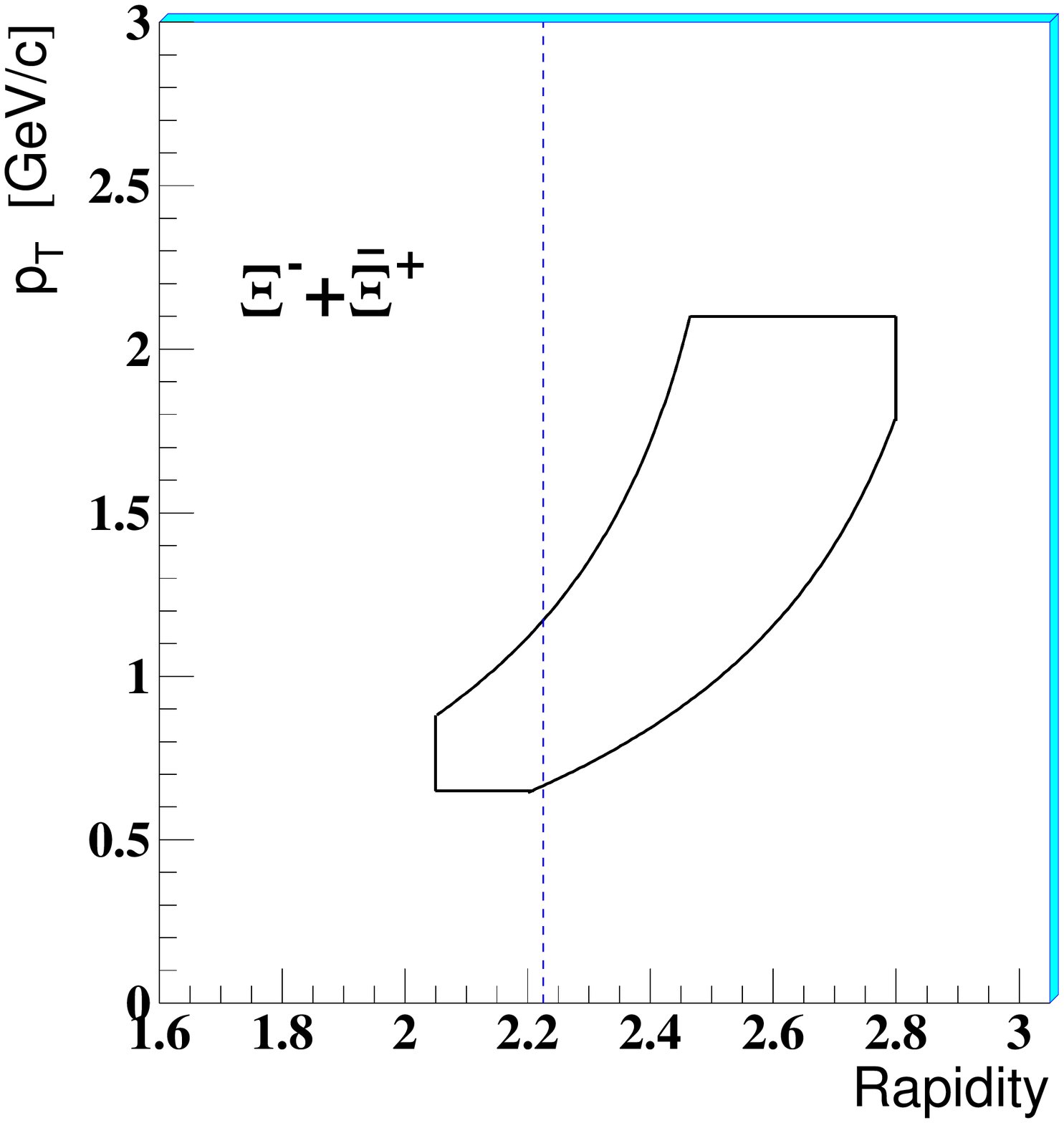}
\includegraphics{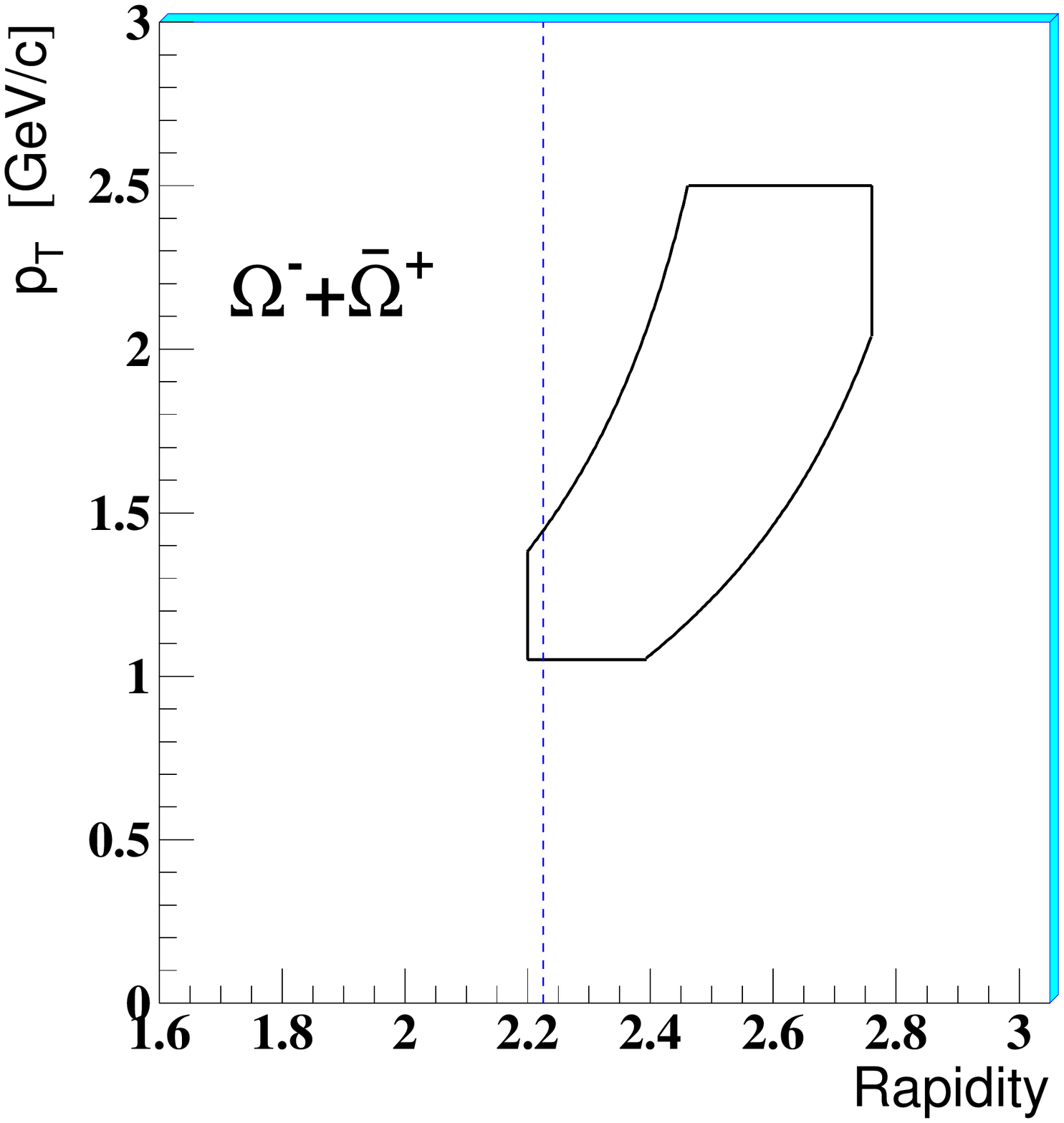}}
\caption{The $y - \pT$ acceptance windows for Pb--Pb collisions. Dashed lines show the value $y_{lab}$ = 2.225, corresponding to the centre of mass mid-rapidity
$\ycm$.}  
\label{fig:accept_pb}
\end{figure}

\begin{figure}[t]
\centering
\resizebox{0.8\textwidth}{!}{%
\includegraphics{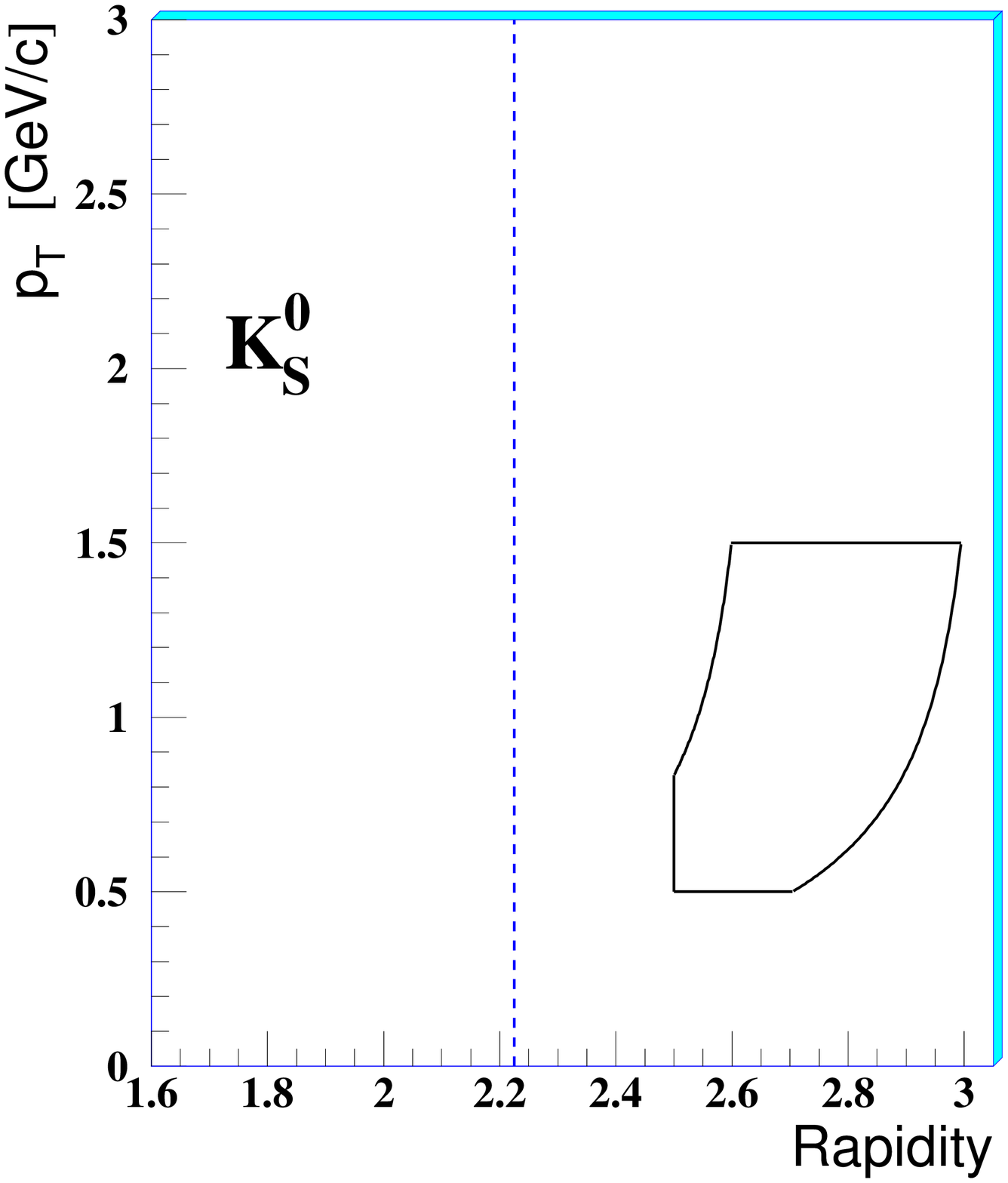}
\includegraphics{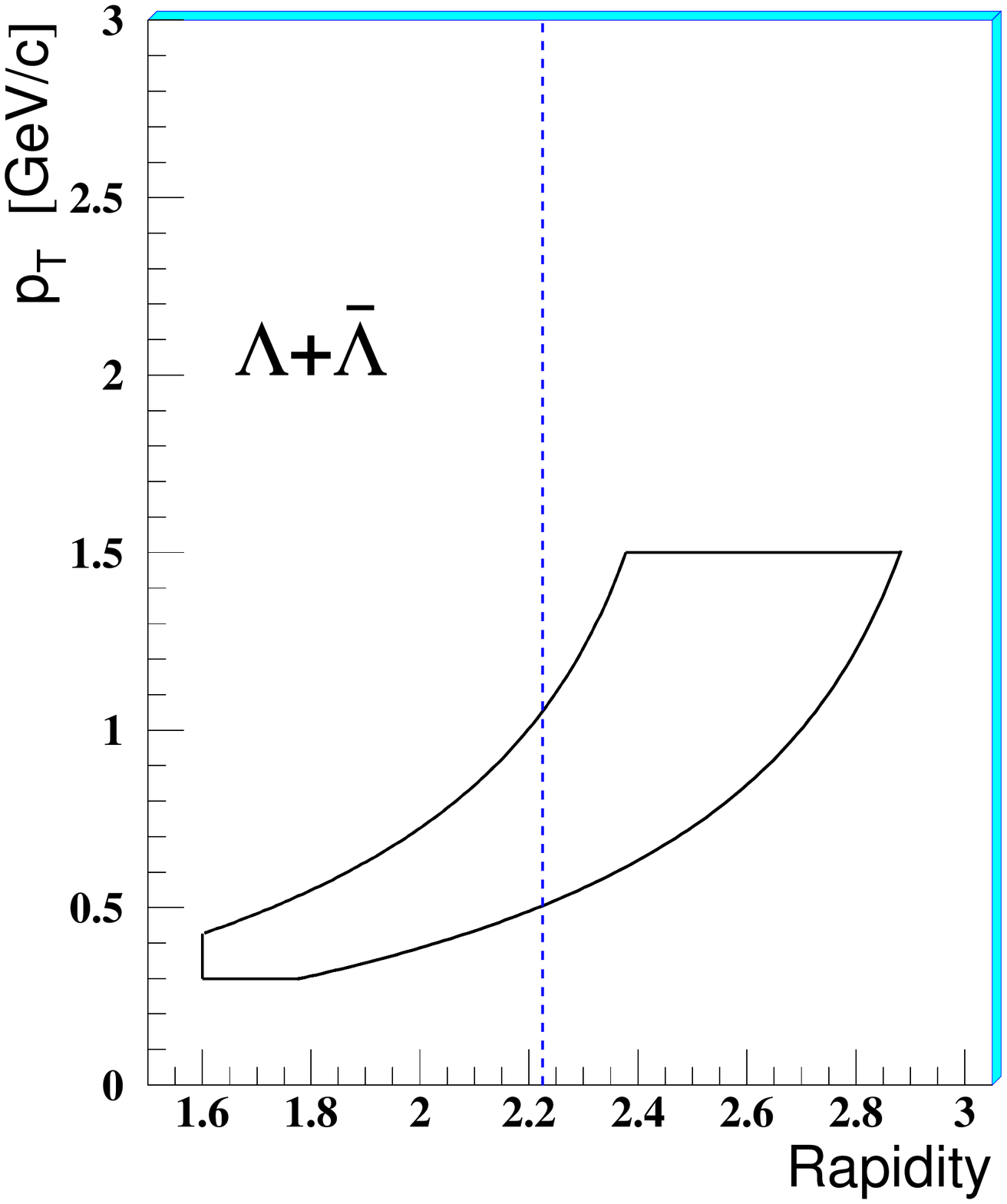}
\includegraphics{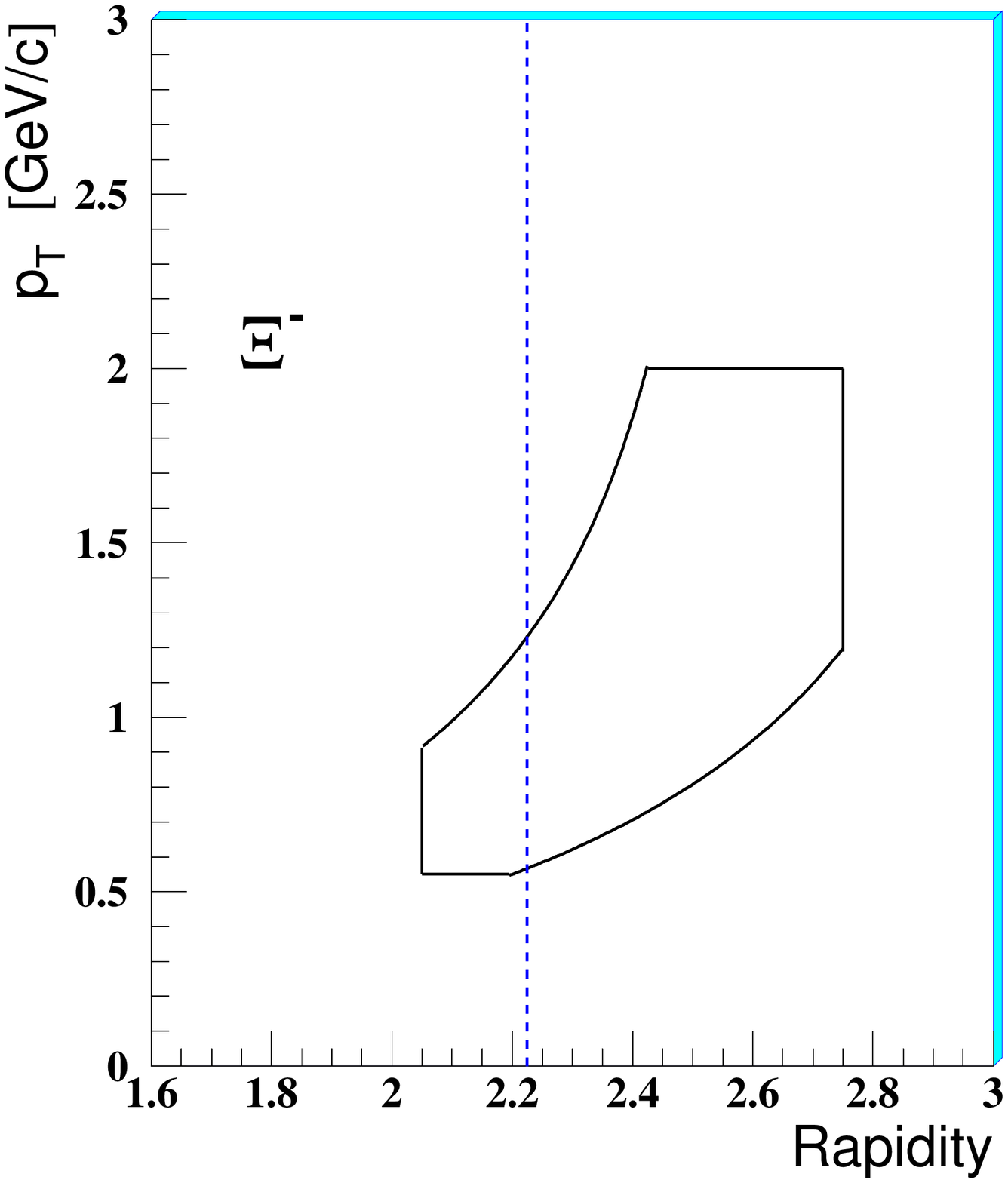}}
\caption{The $y - \pT$ acceptance windows for p--Be collisions. Dashed lines show the value $y_{lab}$ = 2.225, corresponding to the centre of mass mid-rapidity
$\ycm$.}  
\label{fig:accept_be}
\end{figure}

The acceptance windows for $\PKzS$ and hyperons from the Pb--Pb and p--Be collisions are shown in figures \ref{fig:accept_pb} and \ref{fig:accept_be}. The windows have been determined, using a Monte Carlo simulation of the apparatus, requiring 
to exclude from the final sample those particles whose lines of flight are close to the borders of the telescope, where the systematic errors are difficult to evaluate.

The data have been corrected for geometrical acceptance and for reconstruction and signal-extraction losses, using a Monte Carlo weighting procedure on a particle-by-particle basis. The method used was the same as that applied at higher energy \cite{na57_mt1,trans_wa97}. It is based on embedding simulated strange particle decays in real events of similar telescope hit occupancy.  For the p--Be sample the fulfilment of the trigger conditions (at least two charged particles in the SPH1, SPH2 detectors and at least one particle in the ST2 scintillator) was also required.
This requirement must have been fulfilled by the simulated strange decay products in Monte Carlo and was also one of the offline selection criteria for hyperon candidate selection -- each selected candidate had to be able to activate the p--Be trigger. This procedure allowed us to eliminate any influence of the underlying event. The
probability to have more than one strange particle decay fully contained within the NA57 acceptance in p--Be collisions at 40 GeV/c is negligible.

The weighting method described above is CPU intensive, and therefore, for the abundant Pb--Pb samples of $\PKzS,\PgL$ and $\PagL$, we only corrected a fraction of the total data sample in order to reach a statistical precision better than the limits imposed by the systematics. For the p--Be data the weight was calculated for each analysed particle.

The correction procedure has been checked by comparing real and Monte Carlo distributions for a number of parameters, including those used for signal selection (for example the distance of closest approach in space between the two decay particles, the position of the decay vertices) for different data taking periods and magnetic field polarity. 

For the geometrical arrangement of this experiment the feed-down from weak decays
of heavier particles at the given collision energy is negligible. 

As a measure of the collision centrality we use the number of wounded nucleons $N_{\mathrm{wound}}$ \cite{bialas} extracted from the charged particle multiplicity distribution. The procedure to determine the average number of wounded nucleons $\langle N_{\mathrm{wound}} \rangle$ is based on the Glauber model, and has been described in \cite{nicola,centrality}. The Pb--Pb data sample has been divided into five centrality classes (0, 1, 2, 3 and 4 -- class 4 being the most central) according to the value of the charged particle multiplicity measured by the MSD. The fractions of the inelastic cross-section for the five classes are the same as those defined at 158 $A$ GeV/$c$ beam momentum. They are presented, together with the corresponding $\langle N_{\mathrm{wound}} \rangle$ values, in table \ref{tab:centrality}.

\begin{table}[h]
\caption{Centrality ranges and values of
$\langle N_{\mathrm{wound}}\rangle$ for the selected classes\\ in Pb--Pb collisions.
\label{tab:centrality}}
\begin{center}
\begin{tabular}{cccccc}
\hline
 Class &   $0$   &   $1$   &   $2$   &  $3$   &   $4$ \\ \hline
 $\sigma/\sigma_{\mathrm{inel}}$\ \; (\%)   & 40 to 53 & 23 to 40 & 11 to 23 & 4.5 to 11 & 0 to 4.5 \\
 $\langle N_{\mathrm{wound}}\rangle$ & $57 \pm 5$ & $119 \pm 5$ & $208 \pm 4$ & $292 \pm 1$ & $346 \pm 1$ \\ 
 \hline
\end{tabular}  
\end{center}
\end{table}

The number of wounded nucleons in p--Be collisions ($\langle N_{\mathrm{wound}}\rangle$ = 2.5) has also been determined from a Glauber model calculation as an average over all inelastic collisions.

\section{Strange particle yields and enhancements}

The double differential cross-sections for each particle under study were fitted 
using the expression
  \begin{equation}
   \frac{\dder^2 N}{\dder\mT\,\dder y} 
    = f(y)\,\mT\,\exp\left(-\frac{\mT}{T}\right)~,
  \label{ymtdist}
  \end{equation}
where $\mT=\sqrt{m^2+p_{\mathrm{T}}^2}$ is the transverse mass and $y$ is the rapidity. The inverse slope $T$ can be interpreted as an apparent temperature due to the thermal motion coupled with a collective transverse flow of the fireball components \cite{ssh1,ssh2}. The fits were performed using the method of maximum likelihood, considering $T$ as a free parameter.

\begin{figure}[h]
\centering
\resizebox{0.725\textwidth}{!}{%
\includegraphics{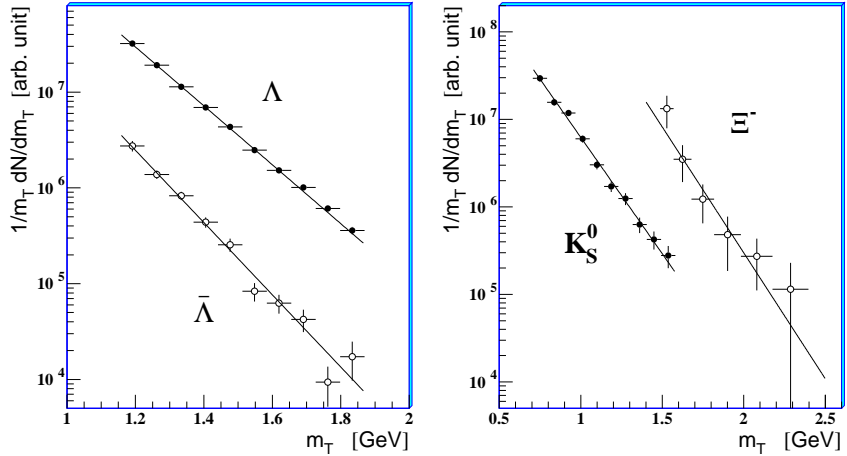}}
\caption{Transverse mass spectra of strange particles from p--Be interactions. The superimposed exponential functions have inverse slopes equal to the $T$ values obtained from the maximum likelihood fits.}  
\label{fig:mt_spectra_be}
\end{figure}

The complete results concerning the transverse mass spectra and their inverse slopes
for strange particles produced in Pb--Pb collisions at 40 $A$ GeV/$c$ are given in \cite{na57_mt2}. Here we present, for the first time, the transverse mass spectra of 
$\PKzS,~\PgL,~\PagL$ and $\PgXm$ from p--Be collisions at the same energy. They are shown in figure \ref{fig:mt_spectra_be}.

The shapes of the distributions are close to exponential and the inverse slopes $T$ extracted from the maximum-likelihood fits are given in table \ref{tab:pbe_data}. In contrast to the Pb--Pb data, the inverse slopes of  $\PgL$ and $\PagL$ spectra differ significantly from each other. Difference between the inverse slopes of the $\PgL$ and $\PagL$ transverse mass spectra has also been observed at 158 $A$ GeV/$c$ \cite{na57_mt1}.

\begin{table}[h]
\caption{Strange particle inverse slopes $T$ and yields $Y$ in p--Be
 collisions.\\ Errors quoted include also a systematic part. \label{tab:pbe_data}}
\footnotesize{
\begin{center}
\begin{tabular}{ccccc}
\hline
\rule[-2.6mm]{0mm}{7.8mm}Particle & $\PKzS$ & $\PgL$ & $\PagL$ & $\PgXm$ \\ 
\hline
 \rule{0mm}{5.5mm}$T$ (MeV) & $169 \pm 6$ & $145 \pm 7$ & $112 \pm 5$ & $151 \pm 28$ \\ 
 \rule[-2.2mm]{0mm}{5.2mm}$Y$   & $0.0354 \pm 0.0030$ & $0.0340 \pm 0.0059$ & 
$0.00255 \pm 0.00023$ & $0.00115 \pm 0.00034$ \\
 \hline
\end{tabular}
\end{center}
}
\end{table}
  
For the present analysis, we assumed a flat rapidity distribution ($f$($y$) = const.)
in our acceptance region for all particles except $\PagL$ and $\PKzS$ in Pb--Pb and $\PKzS$ in p--Be interactions. The rapidity distributions for these two particles are significantly non-uniform and a Gaussian provides a better description.        

Using the parametrization (\ref{ymtdist}), with the value of the
inverse slope $T$ extracted from a maximum likelihood fit to the data, 
we determine the yield $Y$ (number of particles per event) of each particle under study, extrapolated to a common phase-space region covering the full $\mT$ range and one unit of rapidity centred at mid-rapidity $\ycm$ :

\begin{equation}
Y = \int^{\infty}_m \!\!\dder\mT 
\int^{\ycm+0.5}_{\ycm-0.5} \! 
\frac{\dder^2 N}{\dder\mT\,\dder y} ~\dder y~.
\label{ex_yield}  
\end{equation}

Extrapolated yields for p-Be collisions are given in table \ref{tab:pbe_data}. 

For the Pb--Pb interactions the extrapolated yields were determined for each of the five centrality classes and the resulting values are presented in table \ref{tab:yields_pb}.

\begin{table}[h]
\caption{Strange particle yields $Y$ in Pb--Pb for the five centrality classes.\\ (Errors are statistical only).
\label{tab:yields_pb}}
\footnotesize{
\hspace{-0.6cm}
\begin{center}
\begin{tabular}{cccccc}
\hline
  Class &   0      &    1      &    2      &    3    &    4   \\ \hline
\PKzS & $1.28\pm0.17$ & $3.69\pm0.21$ & $7.97\pm0.38$ & $14.06\pm0.68$ &
        $18.38\pm1.10$\\ \hline
\PgL  & $1.55\pm0.13$ & $3.82\pm0.17$ & $9.15\pm0.29$ & $15.24\pm0.53$ & 
        $21.12\pm0.78$\\ \hline
\PagL & $0.044\pm0.006$ &$0.136\pm0.008$ & $0.246\pm0.013$ & 
        $0.350\pm0.019 $ & $0.446\pm0.030$\\ \hline
\PgXm & $0.08\pm0.03$ & $0.31\pm0.05$ & $0.71\pm0.08$ & $1.75\pm0.21$ &
   $1.99\pm0.31$\\ \hline
\PagXp & $0.019\pm0.011$ & $0.025\pm0.016$ & $0.075\pm0.025$ & 
         $0.061\pm0.030$ & $0.079\pm0.031$\\ \hline
$\PgOm+\PagOp$ & & $0.031\pm0.019$ & $0.052\pm0.036$ & $0.106\pm0.056$ &
         $0.144\pm0.083$\\ \hline
\end{tabular}
\end{center}
}
\end{table}

In order to check the stability of the particle yields and $\mT$ spectra the choice of the fiducial regions and of the selection and analysis cuts have been varied. We estimate that the systematic errors on the particle yields and $\mT$ spectra inverse slopes, caused by our selection, correction and analysis procedure, do not exceed 15\% for $\Omega$ and 10\% for all other particles. Independent analyses of two subsamples of p--Be collisions collected in separate running periods allowed for a detailed evaluation of the systematic errors. 

Using the extrapolated yields we have determined the strangeness enhancement $E$, which we define as the yield per participant in Pb--Pb collisions normalized to the yield per participant in p--Be collisions:

\begin{equation}
 E = \left(\frac{Y}{\langle {N_{\mathrm{wound}}} \rangle } \right)_{\mathrm{Pb-Pb}}/
     \left(\frac{Y}{\langle {N_{\mathrm{wound}}} \rangle } \right)_{\mathrm{p-Be}}~.
\label{enhance}  
\end{equation}

\begin{figure}[h!]
\centering
 \resizebox{0.75\textwidth}{!}{%
 \includegraphics{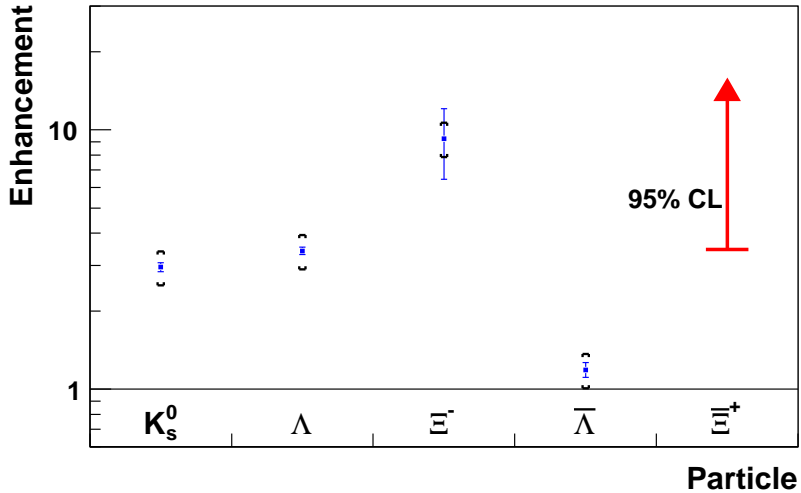}}
\caption {\label{enh_4bins}Strangeness enhancements in Pb--Pb
   collisions at 40 $A$ GeV/$c$. The symbols $^\sqcap_\sqcup$ show the systematic
   errors.}
\end{figure}

The overall enhancements for classes 1 to 4 grouped together (corresponding to the 40\% most central Pb--Pb events) are displayed, as a function of the strangeness content of the particle (anti-particle), in figure \ref{enh_4bins}. For the $\PagXp$, due to the limited statistics of p--Be data, we present only a lower limit to the enhancement at the 95\% confidence level.

A significant enhancement of strangeness production when going from p--Be to \mbox{Pb--Pb} is observed for all strange particles except the $\PagL$. We observe similar enhancement values for the singly strange $\PKzS$ and $\PgL$. The enhancement of the doubly strange $\PgXm$ is larger, reaching about a factor 10. Despite the poor statistics for $\PagXp$ in p--Be data, its production also seems to be enhanced with respect to that of the singly strange $\PagL$.

\begin{figure}[h!]
\centering
 \resizebox{0.7\textwidth}{!}{%
 \includegraphics{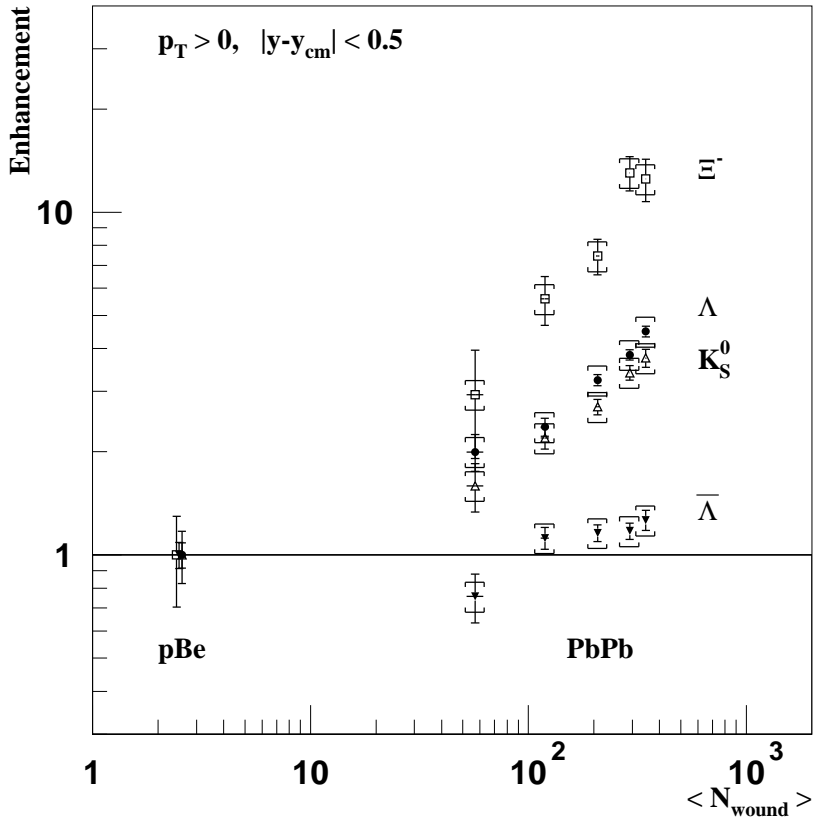}}
\caption {\label{enh_cdep}Strangeness enhancements in Pb--Pb
   collisions at 40 $A$ GeV/$c$ as a function of the collision centrality. 
   The symbols $^\sqcap_\sqcup$ show the systematic errors.}
\end{figure}

The centrality dependence of strangeness enhancements $E$ is shown in figure \ref{enh_cdep}, where the Pb--Pb data, for the five centrality classes defined above, are drawn as a function of the average number of wounded nucleons $\langle N_{\mathrm{wound}}\rangle$.

Significant centrality dependence of the enhancements is observed for all strange particles except for $\PagL$. 

\section{Energy dependence of yields and enhancements}

The energy and centrality dependence of hyperon yields from Pb--Pb collisions
has been studied in the past \cite{elia1,elia2}.

\begin{figure}[h!]
\centering
\resizebox{0.73\textwidth}{!}{%
\includegraphics{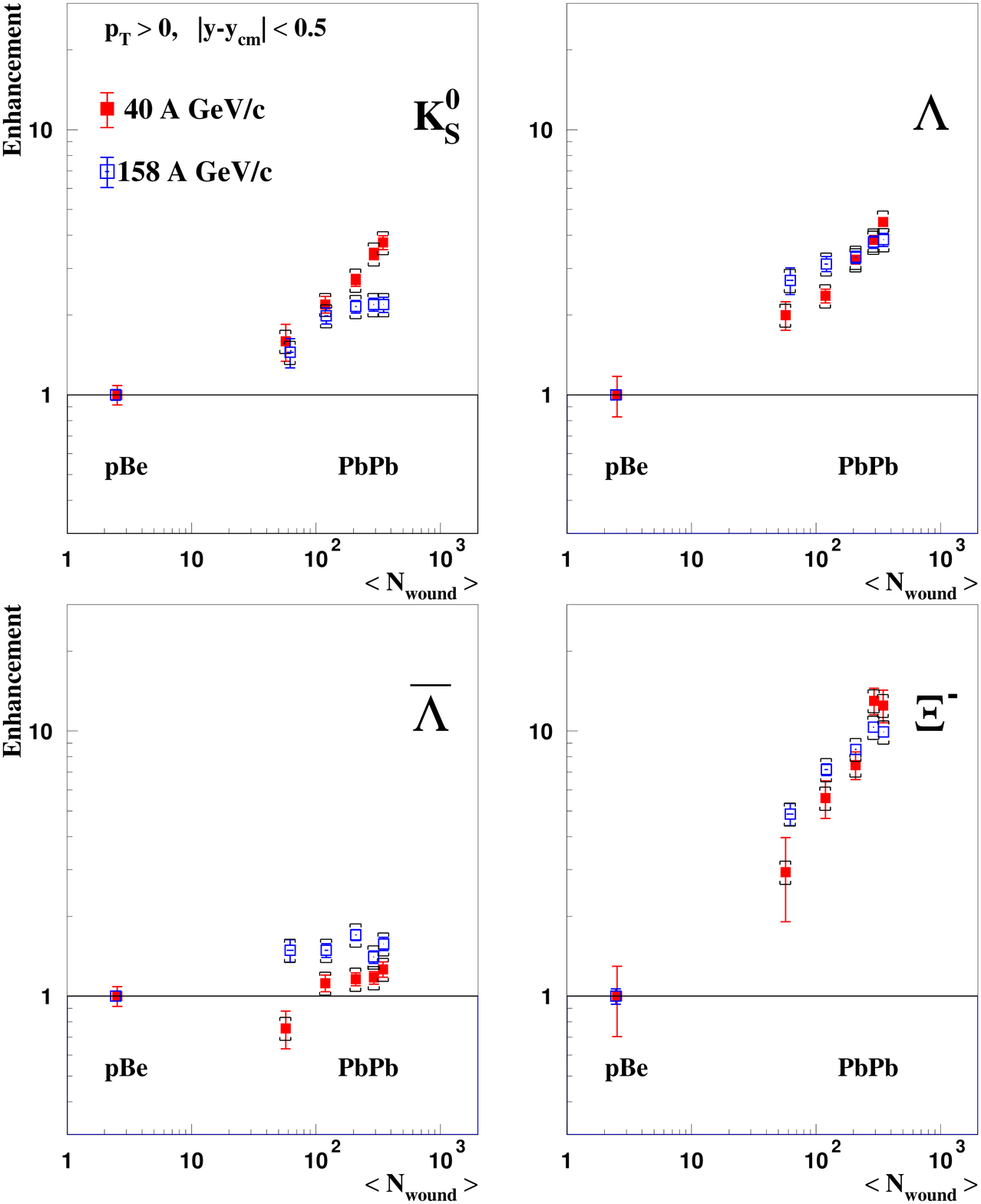}}
\caption{Energy and centrality dependence of strangeness enhancements.}  
\label{fig:enh_edep}
\end{figure}

A comparison of the NA57 Pb--Pb strangeness enhancement measurements at 40 and 158 $A$ GeV/$c$ \cite{enh06} is presented in figure \ref{fig:enh_edep}.

The NA57 results on $\PKzS$ enhancement at 158 $A$ GeV/$c$ are shown here for the first time. The $\PKzS$ p--Be yield $Y = 0.085 \pm 0.004$ at 158 GeV/$c$ has only recently been obtained from an analysis of the WA97 p--Be data sample. Combining this new result with the previously published centrality dependence of the Pb--Pb \mbox{$\PKzS$~yield} \cite{rapidity} allows the corresponding enhancements to be calculated.

For the most central collisions the enhancements for $\PKzS,\PgL$ and $\PgXm$ are higher at 40 than at 158 $A$ GeV/$c$ and a steeper centrality dependence is observed for the lower energy data. The centrality dependence of the $\PKzS$ enhancement at 158 $A$ GeV/$c$ levels off for most central classes. $\PagL$ data do not exhibit a significant enhancement nor centrality dependence at either energy. 

Strangeness enhancements were measured at RHIC by the STAR experiment 
\cite{enh_star,star_overview}. The STAR data show the same enhancement
hierarchy as at the SPS while the magnitudes of the enhancements   
decrease further with the increasing collision energy.

\section{Discussion and conclusions}

We have presented the final NA57 results on strange baryon, anti-baryon and  $\PKzS$
production in Pb--Pb collisions at 40 $A$ GeV/$c$. An enhanced strangeness production
in Pb--Pb with respect to p--Be collisions is observed. The enhancement pattern
follows the same hierarchy as at 158 $A$ GeV/$c$ : i.e. the enhancements increase with the strangeness content of the particle. The magnitude and behaviour of enhancements at different energies is compared. A significant centrality dependence of enhancements is observed at 40 $A$ GeV/$c$ for all particles except the $\PagL$. For central collisions (classes 3 and 4) the enhancement is larger at 40 than at 158 $A$ GeV/$c$.

Work is still underway to find a generally accepted theoretical explanation of the centrality and energy dependence of the observed enhancements.

The hierarchy and energy dependence of the enhancements is in qualitative agreement with the removal of canonical suppression, as expected for a deconfined system with a large correlation volume for strangeness conservation. However, the canonical model formulated in \cite{redlich1,redlich2} reproduces neither the steepness of the observed centrality dependence nor the absolute values of enhancements. The recent introduction of an additional parameter, strangeness correlation radius or chemical off-equilibrium strangeness suppression factor, improves the agreement between statistical canonical model and experimental observations \cite{caines,kraus}.

It has been proposed that the centrality dependence of enhancements can be parametrized as a sum of two terms - one proportional to the number of participants $N_{\mathrm{wound}}$ and the other proportional to the number of binary collisions $N_{\mathrm{bin}}$ \cite{helen}. 
This could mean that the centrality behaviour may depend on the relative importance of soft and semihard collisions.

Recently a two-component geometrical approach has been proposed \cite{becat,werner1,werner2}
distinguishing the dense volume area, referred to as \textit{core}, and low density
peripheral region, referred to as \textit{corona}. The core region participates in 
a collective expansion and produces particles in a statistical manner, whereas the
corona part consists mainly of hadrons produced in elementary nucleon-nucleon
collisions. This approach could explain the centrality dependence of enhancements for different energies and system sizes by variation of the relative importance of core and corona behaviour (see e.g., \cite{becat2,becat3,werner3}).

The growing amount of data on strangeness production in different nuclear systems and over a wide energy range (see e.g., \cite{na49_1,na49_2,takahashi,lamont})  contributes to the better understanding and stimulates a comprehensive theoretical analysis. However, it is worth noting that no conventional hadronic scenario has so far been able to explain experimental observations on strangeness production in high-energy heavy-ion collisions.

\end{document}